# Electrical Impedance Tomography based on Genetic Algorithm

Mingyong Zhou
School of Computer science and Communication Engineering, GXUST
Liuzhou, China
Zed6641@hotmail.com

*Abstract*—In this paper, we applies GA algorithm into Electrical Impedance Tomography (EIT) application. We first outline the EIT problem as an optimization problem and define a target optimization function. Then we show how the GA algorithm as an alternative searching algorithm for solving EIT inverse problem can be applied and used for EIT applications. In fact, a similar neural computation method based on Back Propagation was proposed in reference [2]. In this paper, we explore evolutionary methods such as GA algorithms and compare them to traditional algorithms such Newton Raphson and other linear methods in respect to computation burden.

## I. INTRODUCTION

EIT(Electrical impedance tomography) is a newly and burgeoning image reconstruction technology[1]. It is required that we need to set a circle of electrodes on the surface of the object, then we can reconstruct the interior image when current being injected from the electrode and receiving from other electrodes. Compared with other image reconstruction technology like CT and MRI, the advantage of EIT is budget and non-invasive. Evolutionary and Intelligent algorithms take an important role in computing now, as one of the most popular intelligent algorithms, neural network algorithm is based on logical inference, which is often rely on computers.

Various design and patents were proposed both in system hardware design and software algorithm designs. We in the past years have designed various algorithms to efficiently calculate the inner impedance tomography using both linear methods and Neural computation methods[1][2]. In fact, with the appearance of neuron computation and evolutionary computation methods, there are more options for us to select the algorithm design for more efficient EIT system. In this paper, we applies a GA based algorithm into this EIT algorithm design. First the EIT problem is outlined and described as an inverse optimization problem, and secondly we introduce GA algorithm and define a target optimization function in terms of EIT problem, and last but not least, we propose a algorithm to solve EIT problem using GA method. We also compare the computation burdens among traditional methods[1], our previously proposed neuronal computation method[2] and this method proposed in this paper .

## II. ELECTRICAL IMPEDANCE TOMOGRAPHY PROBLEM

As the technology of EIT researched in 70s, various kind of image reconstruction algorithms have been used.

Inside of magnetic field, magnetic field effect can be ignored due to the quite low permeability in biological tissue. According to the MaxWell theory and Ohm law, we can get the description like this:

$$-\rho^{-1}\nabla\phi = \vec{J} \quad (2\text{-}1)$$

In this equation, ρ represents the distribution function of resistivity, ϕ means potential distribution, J is the function of boundary electric current density.

Due to there is no electric current through interior of biological tissue, so we get this:

$$\nabla \cdot \vec{J} = 0 \quad (2\text{-}2)$$

Combined (1-1) and (1-2), we can get equation :

$$\nabla \cdot \rho^{-1} \nabla \phi = 0 \quad (2\text{-}3)$$

This partial differential equation should satisfy the Dirichlet boundary condition

$$\phi|_{\partial\Omega} = V \quad (2\text{-}4)$$

In that, $\Omega$ is the area where the object in. For the electrodes with inject current, Neumann boundary condition is :

$$\rho^{-1} \frac{\partial \phi}{\partial \vec{n}} = \vec{J} \quad (2\text{-}5)$$

The equations above constitute the mathematical model of EIT

problem.

### III. ALGORITHM DESIGN BASED ON GENETIC ALGORITHM (GA)

In this section, we show how to apply the GA algorithm into EIT inverse optimization problem by defining a optimization function first. We then outline the whole algorithm design details and procedure to solve the EIT problem.

The target optimization function can be defined as in equation (3-1),

$$f = \| y - h(\rho) \|_2 + \alpha \Psi(\rho) \qquad (3\text{-}1)$$

Where $\|.\|_2$ denotes l2 norm in discrete space, $y$ is a column vector whose elements are the measured voltages at surface. $\rho$ is also a vector whose elements are $\rho_1, \rho_2, \ldots \rho_n$. $h$ is the known forward function that calculates the surface voltages given $\rho$, $\alpha, \Psi(\rho)$ are regular parameter and functions that can be classical Tikhonov or Mumford-Shah functions. The EIT inverse problem is to find a set of inner impedance values $\rho$'s so that the function $f$ defined in Equation (3-1) is globally minimized. Hence, we embed the EIT problem as in inverse optimization problem into GA algorithm, and we propose an algorithm as follows:

1), initialize a set of $\rho_1, \rho_2, \ldots \rho_n$ as initial estimate of inner impedance values of the body,
2), define a target optimization function as in Eq.(3-1) to be minimized as a global minimum function as used in GA algorithm.
3), apply the mutation procedure in GA
4), apply the crossover procedure in GA
5), obtain an updated value set for $\rho_1, \rho_2, \ldots \rho_n$..
6), repeat 3)-6) until the updated values $\rho_1, \rho_2, \ldots \rho_n$'s are stable or the value obtained by Equation (3-1) is small enough as required, or the GA generations iterations are expired.

### IV. MATLAB SIMULATIONS FOR GA-EIT

We use 576 inner impedance divisions as shown in Figure 1 and 16 electrodes to measure voltages at surface. In total 576 triangles as shown are used in the simulations. All other settings are the same as in EIDORS v3.9.

Figure 2 is existing NP algorithm result after 1% measurement white noise is added. Figure 2A is the result of first applying NP algorithm and secondly applying GA algorithm in which 1% white measurement noise is added and a disturbance of estimate is added. Figure 3 is known inner impedance tomography.

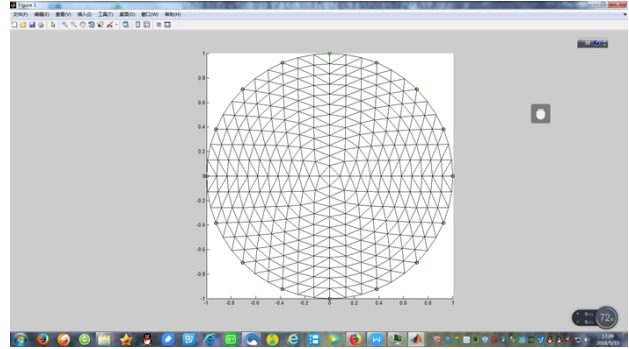

FIGURE 1: TOPOLOGY OF 576 TRIANGLE ELEMENTS TO DIVIDE THE INNER IMPEDANCE

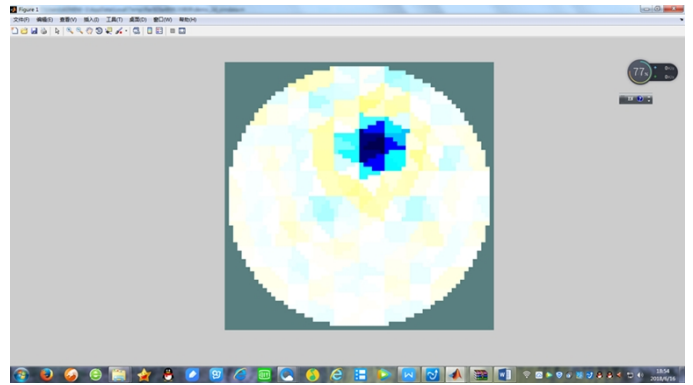

FIGURE 2: SIMULATION RESULT BY NP ALGORITHM

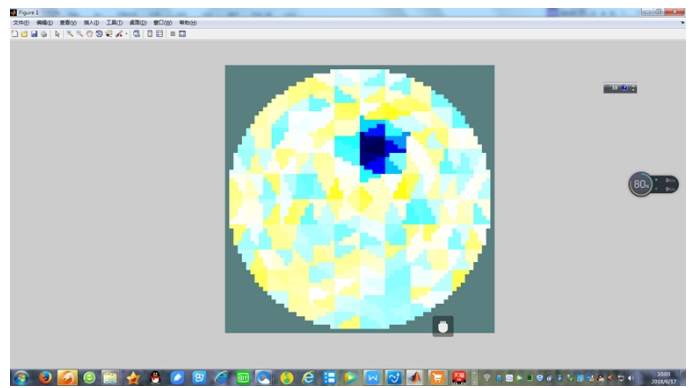

FIGURE 2A: SIMULATION RESULT BY NP & GA WHEN THERE IS 1% MEASUREMENT NOISE AND A DISTURBANCE IS ADDED.

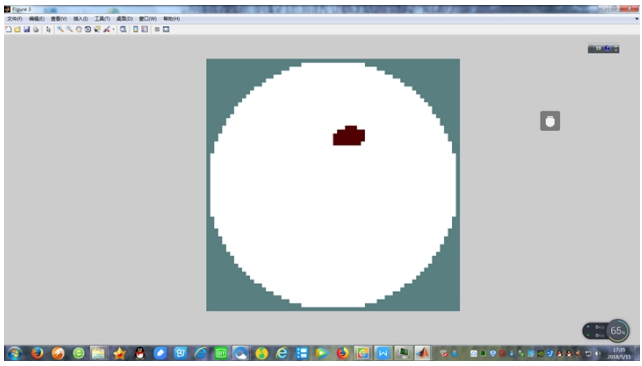

FIGURE 3: ORIGINAL INNER IMPEDANCE TOMOGRAPHY

## V. COMPARISION AMONG GA ALGORITHM, BACK PROPAGATION ALGORITHM AND LEVENBERG ALGORITHM

In reference[1], an algorithm based on Levenberg method is proposed, but the computation involves the design of differential matrix that can increase the computation burden with the increase of inner impedance partitions. A design method is proposed in [1] that could avoid the illness of matrix involved in the calculation of inverse matrix. In reference [2], however a Back Propagation (BP) method based on Neuronal computation is proposed as an alternative method. This method was based on the learning approach used in BP algorithm that requires the neuron network's weights to be learned and decided. In reference [2] design, we assign the impedance value to be associate with each weights in neuron network. A training process must be applied first in order to apply this BP based algorithm into practical use and some simulations were demonstrated in reference[2]

Using EIDORS open platform, we carry out two 2-D simulations and compared the results. Figure 2 is the simulation that is simulated using NP algorithm .Figure 2A is the simulation result using NP+GA algorithm with the default values as in GA function default values of MATLAB 2014a version. Compared with [1], GA based algorithm does not require much matrix operations thus does not need relaxation method to perform matrix calculations. The result may indicate that unconventional methods such as GA-EIT that avoids gradient estimation could perform better in EIT applications. One theoretical explanation is that unconventional methods such as GA-EIT can handle those abruptly changing edges more efficiently without searching the gradients that may either explode or disappear during solution searching process.

## FURTHER CONSIDERATIONS

We are exploring various possibilities to employ other evolutionary methods such as particle swarm optimization(PSO), bat algorithm and neuronal computation without training such as Hopfield's neuron network model to be applied into EIT and check the optimization results. An interesting approach could be that once we can model the EIT problem as a linear matrix approximation, it is possible to exploit the Hopfield neuron model to estimate. A key factor is to measure the optimization time consumed as well as the resolution results that can be gained. Other key factors are the convergence of the proposed algorithms and convergence speed at various scenarios. All these remains as future topics.